\documentclass[showpacs,floatfix,citeautoscript,twocolumn,prb,superscriptaddress]{revtex4-1}

\usepackage{graphicx,xcolor}
\usepackage{amsmath}
\usepackage{comment}
\usepackage{textcomp}
\usepackage{inputenc}

\newcommand*{\cumnas}{Cu$_{0.82}$Mn$_{1.18}$As}

\begin{document}

\title{An in-plane hexagonal antiferromagnet in the Cu-Mn-As system, Cu$_{0.82}$Mn$_{1.18}$As}

\author{Manohar H.\ Karigerasi}
\author{Kisung Kang}
\author{Arun Ramanathan}
\affiliation{Department of Materials Science and Engineering
and Materials Research Laboratory, University of Illinois at Urbana-Champaign, Urbana, IL 61801, USA}
\author{Danielle L.\ Gray}
\affiliation{School of Chemical Sciences, University of Illinois at Urbana-Champaign, Urbana, IL 61801, USA}
\author{Matthias D.\ Frontzek}
\author{Huibo Cao}
\affiliation{Neutron Scattering Division, Oak Ridge National Laboratory, Oak Ridge, TN 37831, USA}
\author{Andr\'{e} Schleife}
\affiliation{Department of Materials Science and Engineering
and Materials Research Laboratory, University of Illinois at Urbana-Champaign, Urbana, IL 61801, USA}
\affiliation{National Center for Supercomputing Applications, University of Illinois at Urbana-Champaign, Urbana, IL 61801, USA}
\author{Daniel P. Shoemaker}\email{dpshoema@illinois.edu}
\affiliation{Department of Materials Science and Engineering
and Materials Research Laboratory, University of Illinois at Urbana-Champaign, Urbana, IL 61801, USA}


\begin{abstract}
We report the single-crystal growth and characterization of a new hexagonal phase, Cu$_{0.82}$Mn$_{1.18}$As, in the Cu-Mn-As system. 
This compound contains the same square-pyramidal MnAs$_5$ units as the tetragonal and orthorhombic polymorphs of CuMnAs.
Calorimetry, magnetometry, and neutron diffraction measurements reveal antiferromagnetic ordering at 270~K.
The magnetic structure consists of a triangular arrangement of spins in the $ab$ plane. 
Hexagonal Cu$_{0.82}$Mn$_{1.18}$As shows resistivity that varies only weakly from 5~K to 300~K, and is many times higher than tetragonal CuMnAs, indicative of a strongly-scattering metal.
First-principles calculations confirm the metallic band structure with a small density of states at the Fermi energy. The neutron-refined
magnetic ground state is close to the computationally-determined minimum energy configuration. This compound should serve as a clear control when disentangling the effects of current-driven N\'{e}el switching of metallic antiferromagnets since it exhibits in-plane spins but the magnetic ordering does not break degeneracy along the $a$ and $b$ directions, unlike tetragonal CuMnAs.
\end{abstract}

\maketitle 

\section{Introduction} 

Recent demonstrations on electronic switching of domains in  semimetallic tetragonal CuMnAs have attracted considerable interest in the field of antiferromagnetic (AF) spintronics.\cite{Wadley2016,Grzybowski2017,Wadley2018,Matalla-Wagner2019} Thin films of tetragonal CuMnAs grown on GaP (001) substrates have a N\'eel temperature $T_N$ of about 480 K.\cite{Wadley2015,Hills2015} 
These studies are complicated by the variable allowed stoichiometries of  phases in the Cu--Mn--As system.
Before any domain-switching studies were demonstrated, bulk tetragonal CuMnAs was shown to be 
stabilized by the addition of excess nominal Cu in solid-state reactions.\cite{Uhlirova2015} 
A large variation in the N\'{e}el temperature $T_N$ from 507~K to 320~K has been shown as Cu excess in Cu$_{1+x}$Mn$_{1-x}$As increases from $x = 0.02$ to 1.4,\cite{Uhlirova2019} and a weak ferromagnetic transition around 300~K was reported around $x=0$.\cite{Nateprov2011}
On the Mn excess side, orthorhombic CuMn$_3$As$_2$ is formed as a stable phase.\cite{Uhlirova2015}
 
When Cu, Mn, and As are mixed stoichiometrically, CuMnAs crystallizes in an orthorhombic $Pnma$ phase \cite{Maca2012}.
Orthorhombic CuMnAs is the first compound to have been proposed as a magnetically-ordered Dirac semimetal \cite{Tang2016} and has been discussed for the possibility of voltage-induced switching \cite{Kim:2018}.
Initial characterization by M\'{a}ca \emph{et al.}\ showed $T_N = 360$~K as judged by resistivity and differential thermal analysis.\cite{Maca2012} This commensurate magnetic ordering and $T_N$ in orthorhombic CuMnAs was confirmed by Emmanouilidou \emph{et al.}, who also found that a slightly cation deficient tetragonal sample Cu$_{0.98}$Mn$_{0.96}$As exhibits an incommensurate AF ordering at 320~K, followed by another AF reorientation around 230~K.\cite{emmanouilidou_magnetic_2017}

\begin{figure}
\centering\includegraphics[width=0.8\columnwidth]{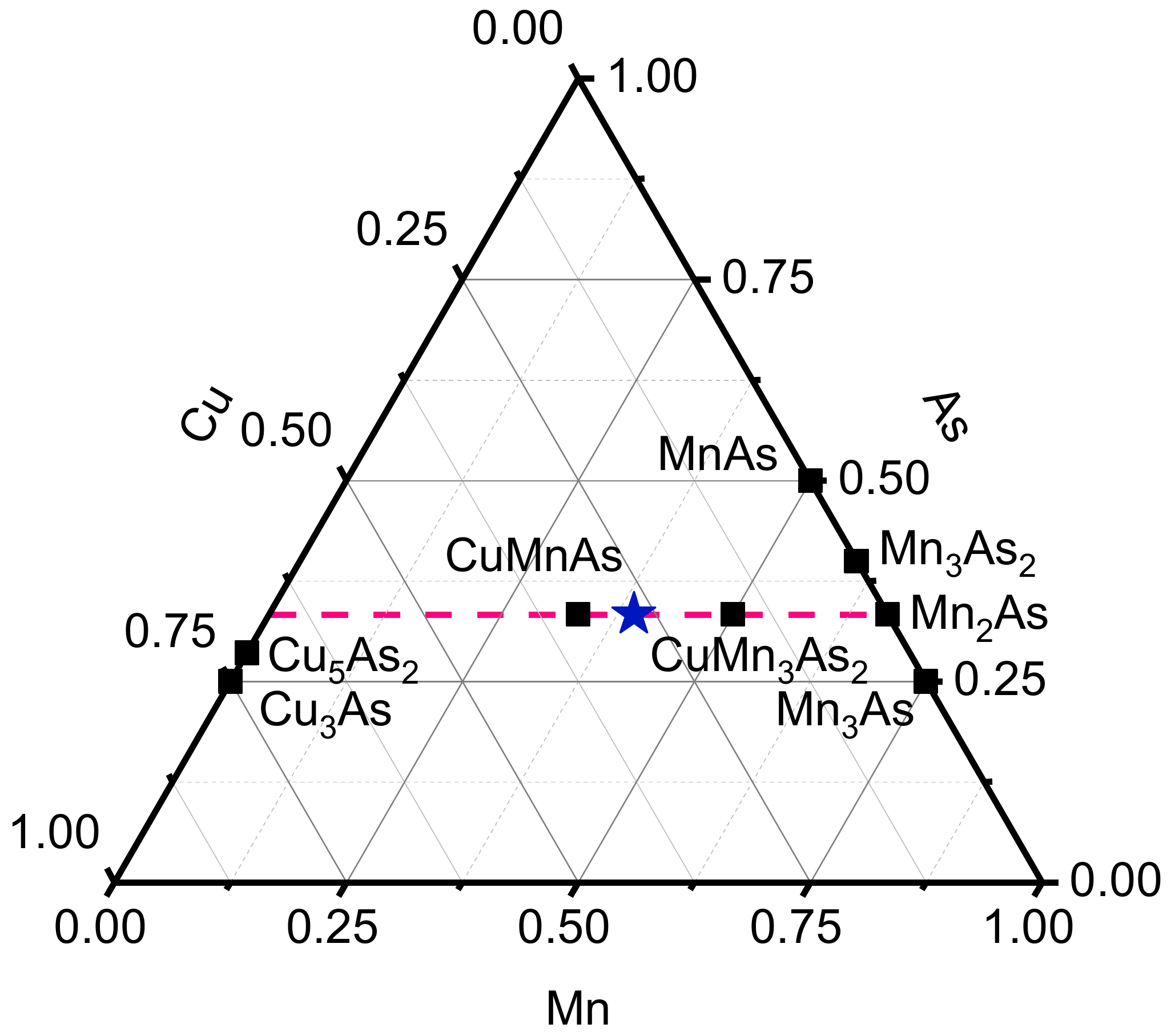} \\
\caption{\label{fig:phase_diagram}(Color online.)
{\color{black}Hexagonal \cumnas\ has been marked with a star among the previously-known phases in the Cu--Mn--As system. Compositions near CuMnAs are known to crystallize in both tetragonal and orthorhombic crystal systems.}
} 
\end{figure}

{\color{black} All known ternary phases in the Cu--Mn--As system have the transition metal ($M$) to As ratio of 2:1 and are either tetragonal or orthorhombic as shown in Fig.\ \ref{fig:phase_diagram}.}
The metallic nature of these compounds allows significant deviation from $M_2$As stoichiometry, as evidenced by the binary compounds MnAs (a ferromagnet with a reentrant FeP-to-NiAs-type transition),\cite{pytlik_magnetic_1985,schwartz_magnetic_1971,Glazkov2003} Mn$_3$As$_2$ (which has at least three polymorphs),\cite{dietrich_crystal_1990,moller_crystal_1993,hagedorn_crystal_1994} the seemingly metastable compounds Mn$_4$As$_3$ and Mn$_5$As$_4$,\cite{hagedorn_synthesis_1995,moller_crystal_1993} and Mn$_3$As.\cite{nowotny_kristallchemische_1951}
Of these compounds, only MnAs and Mn$_2$As have been investigated with neutron diffraction and transport measurements.\cite{yuzuri_magnetic_1960,austin_magnetic_1962}
Further elaboration of compounds in this space is necessary to understand the potential for manipulating spins in these highly-correlated phases.

\section{Methods}

Millimeter-sized crystals of hexagonal \cumnas\ were synthesized by mixing elemental powders Cu (99.9\% metals basis), Mn (99.98\% metals basis), and As (99.9999\% metals basis) in 0.82:1.18:1 molar ratio. 
The powders were vacuum sealed in quartz tubes and heated at 1$^\circ$C/min to 600$^{\circ}$C for 6 hours then ramped at 1$^{\circ}$C/min to 975$^{\circ}$C for 1 hour. The tube was slow cooled at 1$^{\circ}$C/min to 900$^{\circ}$C and held for 1 hour before furnace-cooling down to room temperature. The resulting product was a solid ingot. The ingot was crushed into smaller pieces to conduct single crystal X-ray diffraction on a Bruker X8 Apex II diffractometer at 296~K and $\lambda = 0.71073$~\AA.

Variable-temperature powder X-ray diffraction was performed using a nitrogen blower at beamline 11-BM of the Advanced Photon Source in Argonne National Laboratory ($\lambda = 0.4128$~\AA).\cite{wang_dedicated_2008} Variable-temperature neutron powder diffraction was conducted at the WAND$^2$ instrument at the High-Flux Isotope Reactor (HFIR) at Oak Ridge National Laboratory.\cite{Frontzek_new}

Magnetic structure determination was performed on a 2~mm crystal at the HB-3A four circle diffractometer at HFIR. A total of 344 reflections were collected at 4~K and used for structural refinement. Magnetic 
symmetry analysis was carried out using the tools
available at the Bilbao Crystallographic Server\cite{perez-mato_symmetry-based_2015}
and refined using the FullProf suite.\cite{rodriguez-carvajal_recent_1993}

Differential scanning calorimetry (DSC) measurements were performed on 5~mg of powder in Al pans under N$_2$ atmosphere in a TA Instruments DSC 2500.
A small fractured sample, weighing about 12 mg, was polished and aligned using Laue diffraction. This sample was mounted onto a quartz paddle sample holder for aligned magnetometry measurements in a Quantum Design MPMS3. 
Aligned resistivity measurements were carried out using the 4-point probe method in a Quantum Design PPMS DynaCool.

First-principles density functional theory (DFT) simulations were performed using the Vienna \emph{Ab-Initio} Simulation Package  (VASP) \cite{Kresse:1996,Kresse:1999}.
The electron-ion interaction is described using the projector-augmented wave (PAW) scheme.\cite{Blochl:1994}
Exchange and correlation are described using the generalized-gradient approximation (GGA) by Perdew, Burke, and Ernzerhof  (PBE).\cite{Perdew:1997}
Single-particle Kohn-Sham states are expanded into a plane-wave basis with a cutoff energy of 600 eV.
Monkhorst-Pack \cite{Monkhorst:1976} (MP) $\mathbf{k}$-point grids of $2\,\times\,2\,\times\,6$ and $4\,\times\,4\,\times\,12$ are used to integrate the Brillouin zone for cell relaxation and electronic band structure calculations, respectively.
Non-collinear magnetism and spin-orbit coupling is taken into account in all calculations \cite{Steiner:2016}.
Self-consistent total-energy convergence was achieved to within $10^{-6}$ eV and atomic positions were relaxed until Hellman-Feynman forces were smaller than 5~meV/\AA.

\section{Results and Discussion}

\subsection{Structure refinement}

\begin{figure}
\centering\includegraphics[width=\columnwidth]{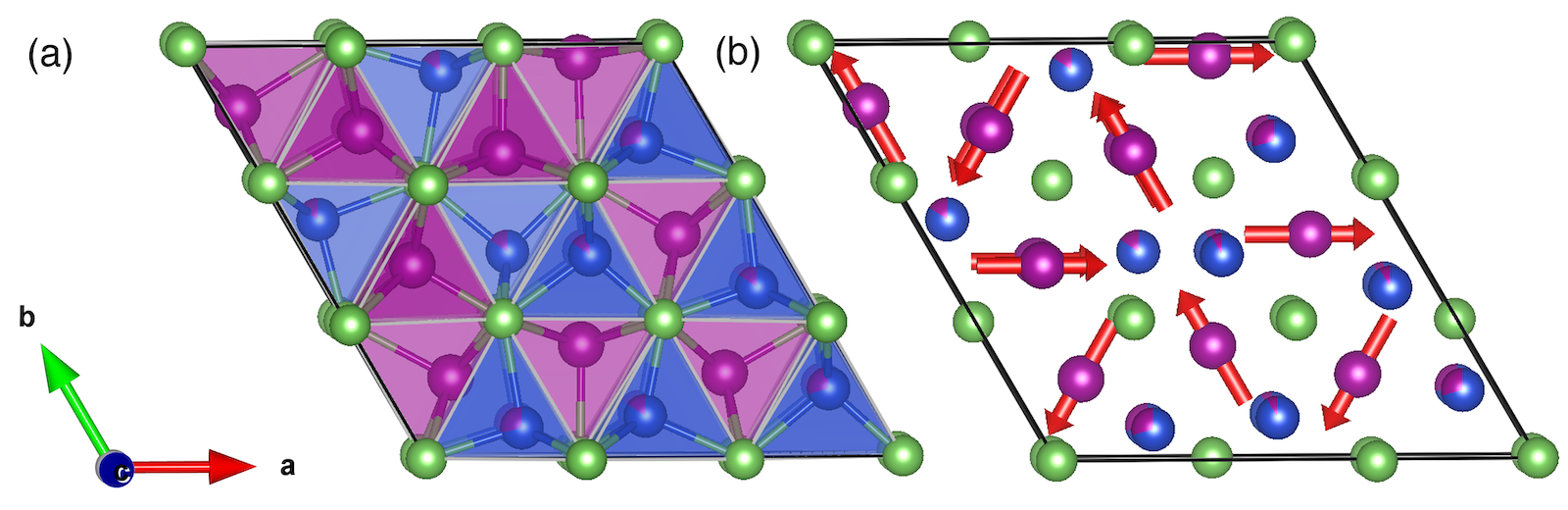} \\
\caption{\label{fig:unitcell}(Color online.)
Unit cell of \cumnas\ (a) is shown with square pyramidal Mn in purple, tetrahedrally coordinated Cu in blue, and As in green.
In (b), the refined magnetic structure is shown with moments on the Mn sites.
} 
\end{figure}

\begin{figure}
\centering\includegraphics[width=\columnwidth]{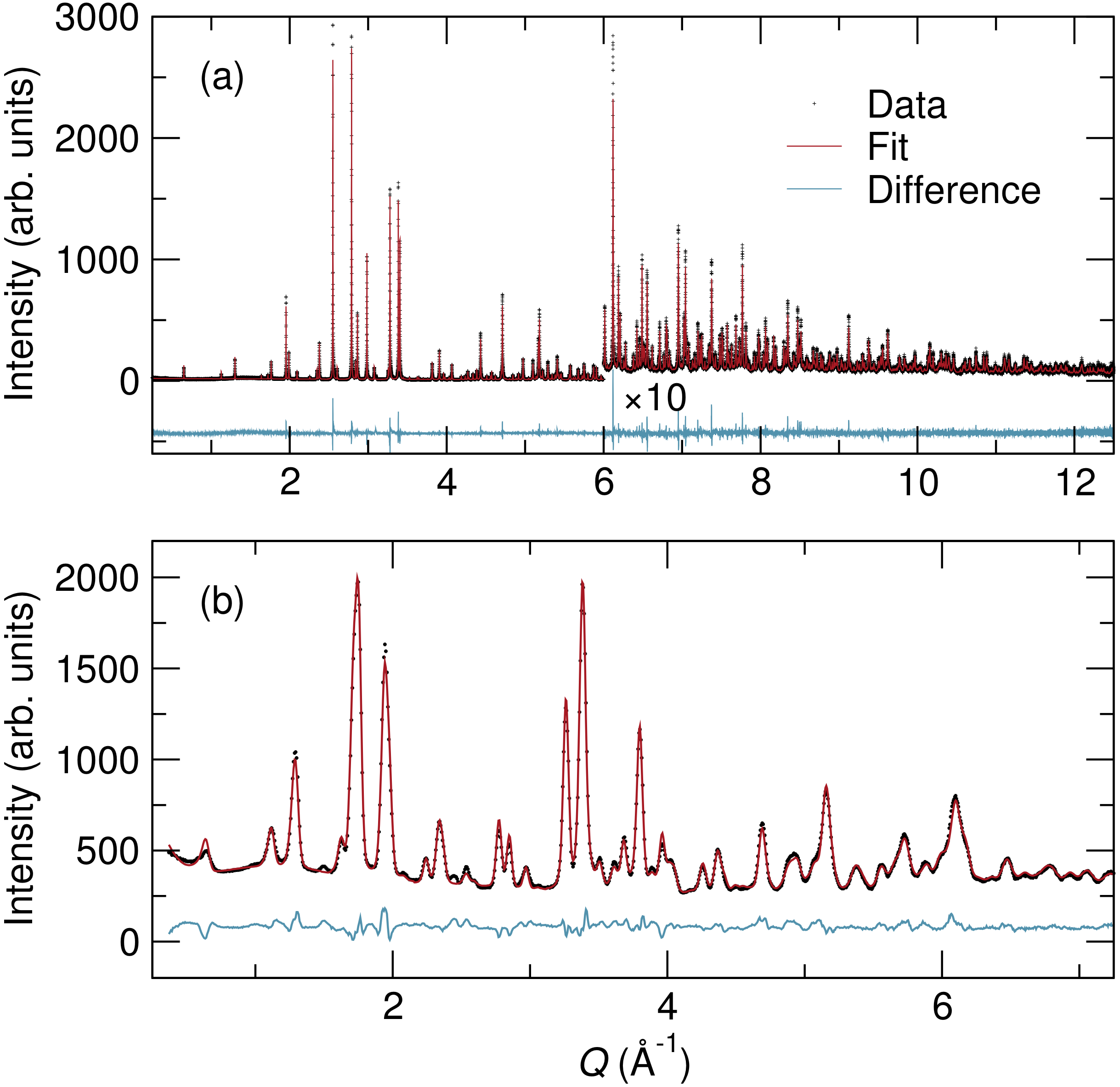} \\
\caption{(Color online.)
Refinements of \cumnas\ to (a) synchrotron X-ray powder diffraction at 100~K at APS 11-BM ($\lambda = 0.4128$~\AA) and (b) neutron powder diffraction at 400~K on WAND$^2$ ($\lambda = 1.487$~\AA).
}
\label{fig:xrd-neutron}
\end{figure}

The refined structure of \cumnas\ is shown in Fig.\ \ref{fig:unitcell}(a), with structural parameters from single-crystal X-ray diffraction (XRD) given in Table \ref{tab:sxtl} and \ref{tab:atoms}.
{\color{black}\cumnas\ has a short lattice parameter $c \approx 3.8$~\AA, indicating that the unit cell is flat and is the same width as the Cu and Mn coordination polyhedra. Fig.\ \ref{fig:unitcell}(a) shows the unit cell viewed down $c$, with all the atoms occupying either $z$ = 0 or $z$ = 0.5.}
The compound forms in a new structure type with space group $P\overline{6}$, and is comprised of three inequivalent square-pyramidal Mn and three inequivalent tetrahedral Cu, all coordinated by As.
All metal sites have a multiplicity of 3 {\color{black} and have $m$ point symmetry}. The atomic positions are well-described by the single-crystal XRD data, but the occupancies are less reliable due to the similar electron densities at each site.  
High-resolution synchrotron powder X-ray diffraction is shown in Fig.\ \ref{fig:xrd-neutron}(a), to confirm that these samples can be made highly pure with excellent crystallinity.

The occupancies are better constrained by neutron scattering, where Mn and Cu have more contrast in their scattering lengths ($-3.73$ and 7.718 fm, respectively).\cite{sears_neutron_1992}
Neutron powder diffraction data from WAND$^2$ were collected at 400~K, in the paramagnetic regime, with the refinement shown in Fig.\ \ref{fig:xrd-neutron}(b). 
No evidence for site mixing or vacancies on the Mn or As sites was apparent. 
The best refinements were obtained by using the nominal Cu/Mn ratio and allowing Mn mixing on the Cu sites, with the final Cu occupancies of 0.709(2), 0.914(3), and 0.846(2) for Cu sites 1\,--\,3, respectively.
The final structural refinement data presented in Table \ref{tab:atoms} is a single-crystal XRD refinement with the occupancies locked to values obtained by co-refinement to the 100~K synchrotron X-ray and 400~K neutron scattering data. 

\begin{table}
\caption{\label{tab:sxtl} 
Structural parameters obtained from room-temperature Mo-$K\alpha$ X-ray
single-crystal refinement (full-matrix least-squares on $F^2$) with occupancies fixed from synchrotron X-ray and neutron co-refinement. 
}
\centering
\begin{tabular}{ll}
\hline
Formula								&  \cumnas\ \\
Formula Weight						& 191.88 g/mol\\
Crystal system						& Hexagonal \\
Space group							& $P\overline{6}$\\
$a=b$									& 11.1418(3) \AA \\
$c$									& 3.8311(2) \AA \\
$V$, $Z$ 								& 411.87(3) \AA$^3$, 9 \\
$\rho$ 								& 6.962 g/cm$^3$ \\
Absorption coefficient			& 35.046 mm$^{-1}$ \\
$F$(000) 							& 777 \\ 
$(\sin\theta/\lambda)_{max}$ 					& 0.714 \\
Reflections collected       	& 6722 \\
Observed $I>2\sigma(I)$ reflections       	& 953 \\
$R_{int}$							& 0.0682 \\
Number of parameters 			& 56 \\
Goodness-of-fit on $F^2$		& 1.445 \\
$R[F^2 > 2\sigma(F^2)], wR(F^2)$    &   0.0344, 0.0849 \\
\hline
\end{tabular}
~\\
\end{table}

\begin{table*}
\caption{\label{tab:atoms} 
Atomic parameters obtained from room-temperature X-ray single-crystal refinement of \cumnas.
Occupancy values for Cu/Mn sites are co-refined to 100~K synchrotron and 400~K neutron powder diffraction data (see Fig.\ \ref{fig:xrd-neutron}).
Atomic displacement parameters $U_{ij}$ are given in units of \AA$^2$.
}
\centering
\begin{tabular}{p{1.7cm}p{0.9cm}p{1.7cm}p{1.7cm}p{0.9cm}p{2.3cm}p{1.7cm}p{1.7cm}p{1.7cm}p{1.7cm}}
\hline\hline
Atom	 & Site	 & $x$	 & $y$	 & $z$	 & Occupancy	 & $U_{11}$	 & $U_{22}$	 & $U_{33}$	 & $U_{12}$\\
\hline
Cu1/Mn1	 & 3j	 & 0.2489(9)	 & 0.0868(10)	 & 0	 & 0.709/0.291(2)	 & 0.025(3)	 & 0.023(4)	 & 0.030(3)	 & 0.014(2)\\
Cu2/Mn2	 & 3j	 & 0.5894(9)	 & 0.5043(5)	 & 0	 & 0.914/0.086(3)	 & 0.012(3)	 & 0.011(2)	 & 0.015(2)	 & 0.005(3)\\
Cu3/Mn3	 & 3k	 & 0.4206(9)	 & 0.5030(6)	 & 0.5	 & 0.846/0.154(2)	 & 0.016(3)	 & 0.012(3)	 & 0.018(3)	 & 0.009(2)\\
Mn4	 & 3j	 & 0.1966(10)	 & 0.4682(10)	 & 0	 & 1	 & 0.016(4)	 & 0.017(4)	 & 0.009(3)	 & 0.011(3)\\
Mn5	 & 3k	 & 0.8051(10)	 & 0.9422(7)	 & 0.5	 & 1	 & 0.015(3)	 & 0.017(2)	 & 0.017(3)	 & 0.009(3)\\
Mn6	 & 3k	 & 0.8056(10)	 & 0.5310(10)	 & 0.5	 & 1	 & 0.009(3)	 & 0.013(3)	 & 0.011(3)	 & 0.006(2)\\
As1	 & 3j	 & 0.3310(6)	 & 0.3354(6)	 & 0	 & 1	 & 0.010(2)	 & 0.008(3)	 & 0.013(3)	 & 0.0044(19)\\
As2	 & 3k	 & 0.6754(6)	 & 0.6721(6)	 & 0.5	 & 1	 & 0.010(2)	 & 0.013(3)	 & 0.0073(14)	 & 0.007(2)\\
As3	 & 1a	 & 0	 & 0	 & 0	 & 1	 & 0.011(2)	 & 0.011(2)	 & 0.007(4)	 & 0.0054(12)\\
As4	 & 1d	 & 1/3	 & 2/3	 & 0.5	 & 1	 & 0.007(2)	 & 0.007(2)	 & 0.012(4)	 & 0.0033(12)\\
As5	 & 1e	 & 2/3	 & 1/3	 & 0	 & 1	 & 0.009(3)	 & 0.009(3)	 & 0.002(4)	 & 0.0045(14) \\
\hline \hline
\end{tabular}
~\\
\end{table*}

\subsection{Magnetic ordering}

In light of the strong exchange coupling in transition-metal arsenides that leads to high Curie and N\'{e}el temperatures in MnAs and Mn$_2$As, it is surprising that few hexagonal arsenides have been shown to order magnetically near room temperature. The most well-known structure type is the $P6_3/mmc$  NiAs-type, of which MnAs is a member. NiAs itself is a Pauli paramagnet,\cite{nozue_specific_1997} while hexagonal CrNiAs has a Curie temperature of 190~K.\cite{stadnik_magnetization_2008}

\begin{figure}
\centering\includegraphics[width=\columnwidth]{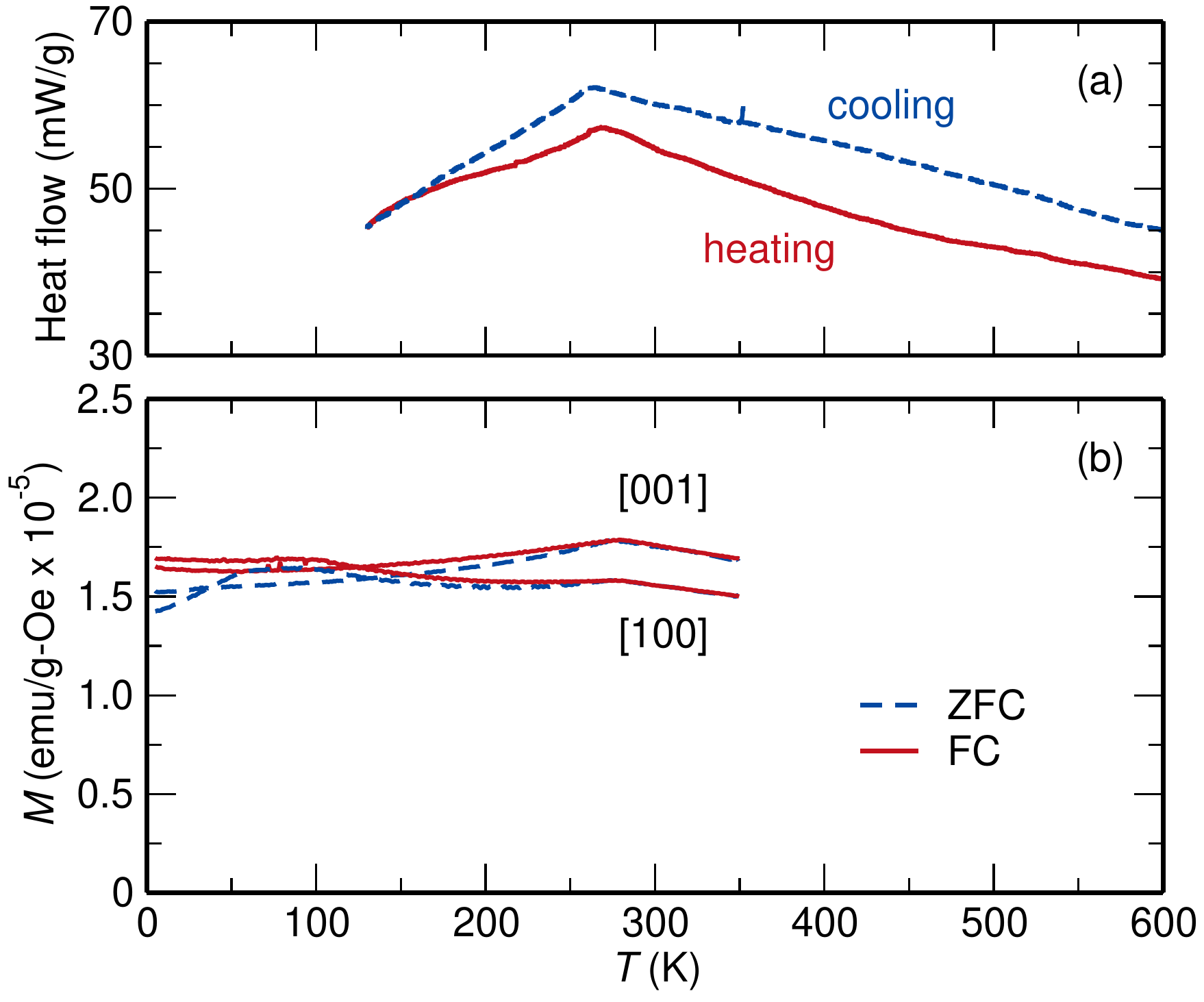}
\caption{(Color online.)
DSC data (a) show a clear kink in the heat flow at $T_N \approx 270$ K, indicating a discontinuous change in heat capacity of the sample. Data on heating are reflected about the $x$-axis. The same transition appears in magnetic susceptibility measurements (b) of an aligned single crystal, with the field axis along the [001] and [100] directions.
} 
\label{fig:dsc-mpms}
\end{figure}

Powders of \cumnas\ were examined by DSC, with the heating and cooling traces shown in Fig.\ \ref{fig:dsc-mpms}(a).
There is a clear change in slope around 267~K, with a hysteresis of about 4~K.
To determine the origin of this transition, aligned single crystals of \cumnas\ were examined via SQUID magnetometry, and the moment versus temperature is shown in Fig.\ \ref{fig:dsc-mpms}(b).
The maximum in the magnetometry data is around 275~K for zero-field-cooled (ZFC) and field-cooled (FC) data along the $a$ and $c$ axes {\color{black} for 10 kOe applied field}. There are not sufficient data above $T_N$ to provide a satisfactory Curie-Weiss fit.
The data along the $c$ axis display a typical decrease upon cooling past $T_N$, while the data measured along the $a$ axis show a slight rise and plateau around 100~K.
There were no features in measurements of magnetic moment versus field to indicate spin-flop transitions or any hysteresis. The small plateau could arise from decreasing itineracy and a leveling-off of the local moments on Mn sites, which would be consistent with the single-crystal neutron magnetic intensity remaining constant below 100~K. 
{\color{black}  Fig. \ref{fig:dsc-mpms}(b) shows that for temperatures beyond 150 K, the susceptibility along $c$ is larger than along $a$. This trend is consistent with in-plane moments in triangular antiferromagnets such as CsMnBr$_3$, CsVCl$_3$, Mn$_3$Sn etc.\cite{Brown_1990,Kotyuzhanskii1991,Hirakawa1983,Duan2015} Below 150 K, the difference in the susceptibility along $a$ and $c$ is unclear. However, the in-plane Mn spin ordering was confirmed using neutron diffraction as shown below.}



\begin{figure}
\centering\includegraphics[width=\columnwidth]{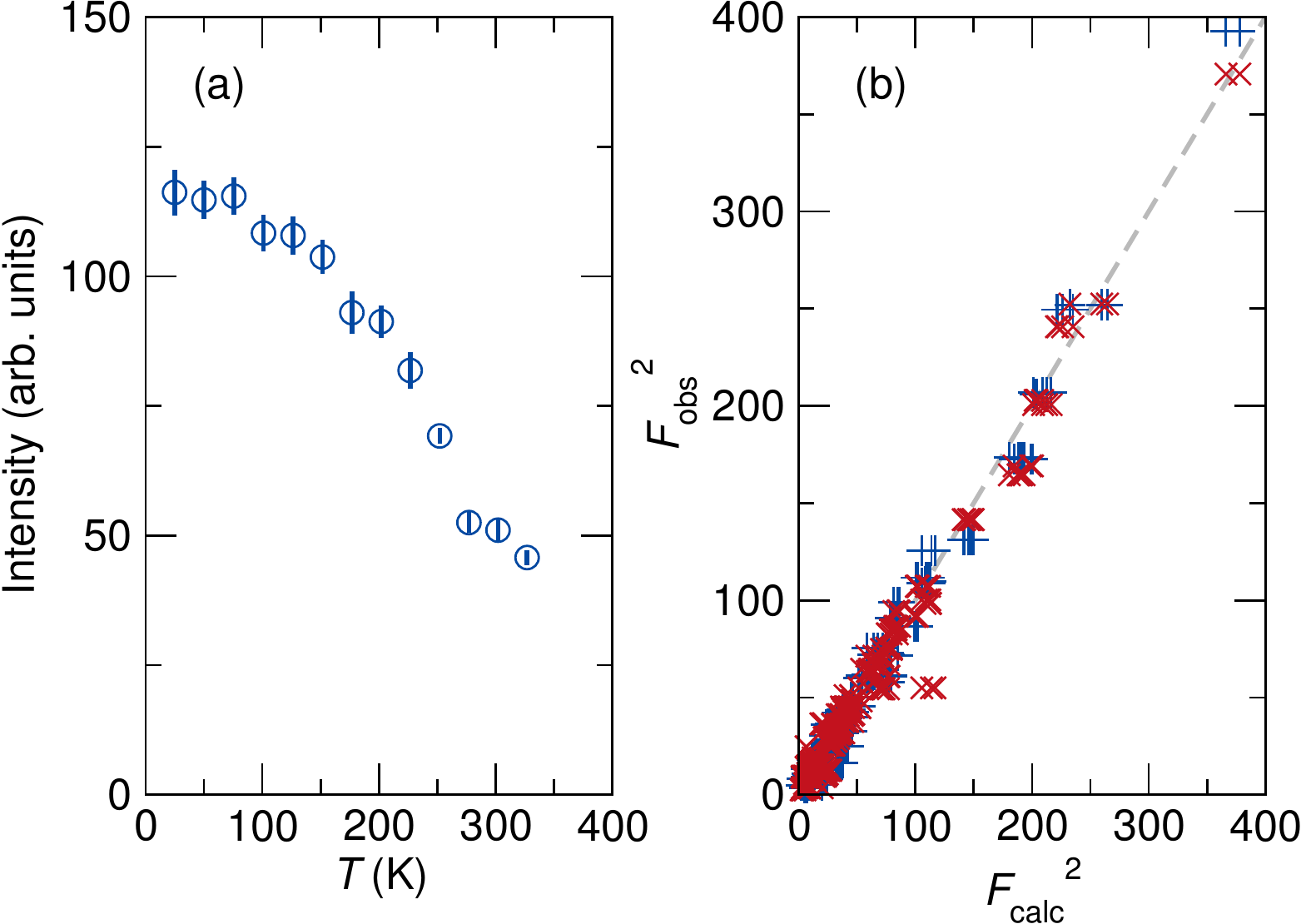} \\
\caption{\label{fig:hb3a}(Color online.)
Measured single-crystal neutron diffraction intensity (a) of the (020) peak of \cumnas\ shows a gradual increase upon cooling past $T_N$ down to 4~K. The (020) peak is an allowed nuclear reflection and persists with constant intensity ($\sim 50$) above $T_N$. The differences between observed and refined structure factors $F_{hkl}^2$ at $T=4$~K are shown in (b). The triangular model obtained from neutron refinement are shown as ($+$) and the DFT-derived model as ($\times$).
} 
\end{figure}

The magnetic ordering was probed first by variable-temperature neutron powder diffraction
on the WAND$^2$ instrument, which showed changes in peak intensities across this boundary, but no new peaks, indicating likely $k = 0$ ordering.
A full triple-axis data collection was performed on the HB-3A beamline at 4~K.
The magnetic and nuclear structures were refined together in the $P\overline{6}^\prime$ magnetic space group. 
The intensity of the (020) peak can serve as an order parameter, and its temperature dependence is shown in Fig.\ \ref{fig:hb3a}(a).
The (020) peak is an allowed nuclear reflection, so the intensity does not go to zero above $T_N$.
The three inequivalent Mn sites are constrained to have equal magnetic moments, which are refined to 3.02(8) $\mu_B$/atom.
No improvement in the fit was observed when the moments were allowed to freely vary.
The observed and calculated structure factors $F_{hkl}^2$ are plotted in Fig.\ \ref{fig:hb3a}(b).
The magnetic structure is shown in Fig.\ \ref{fig:unitcell}(b). 
No local Mn moment was stably refined on the Cu-majority sites, and Cu itself does not host local moments in arsenides \cite{pauwels_electrical_1973,sampathkumaran_enhanced_2003,sengupta_magnetic_2005}.
It is possible that some local Mn moments exist on the minority Cu sites, but they do not appear to be ordered.
{\color{black}The 120$^\circ$ spin structure differs from Mn$_3$Sn. In Mn$_3$Sn, the spin triangles are connected by their corners. We also do not observe the ``inverse triangle'' orthorhombic configuration seen in Mn$_3$Sn.\cite{Brown_1990} There are three different types of 120$^\circ$ spin structures observed in our compound, although the spin directions in the $ab$ plane could not be uniquely determined by unpolarized neutron diffraction.}
This compound could be written as containing Cu$^+$ and Mn$^{2+}$, but like other transition-metal arsenides the local moment is reduced due to metallicity.\cite{katsuraki_magnetic_1966,pytlik_magnetic_1985}

\begin{figure}
\centering\includegraphics[width=\columnwidth]{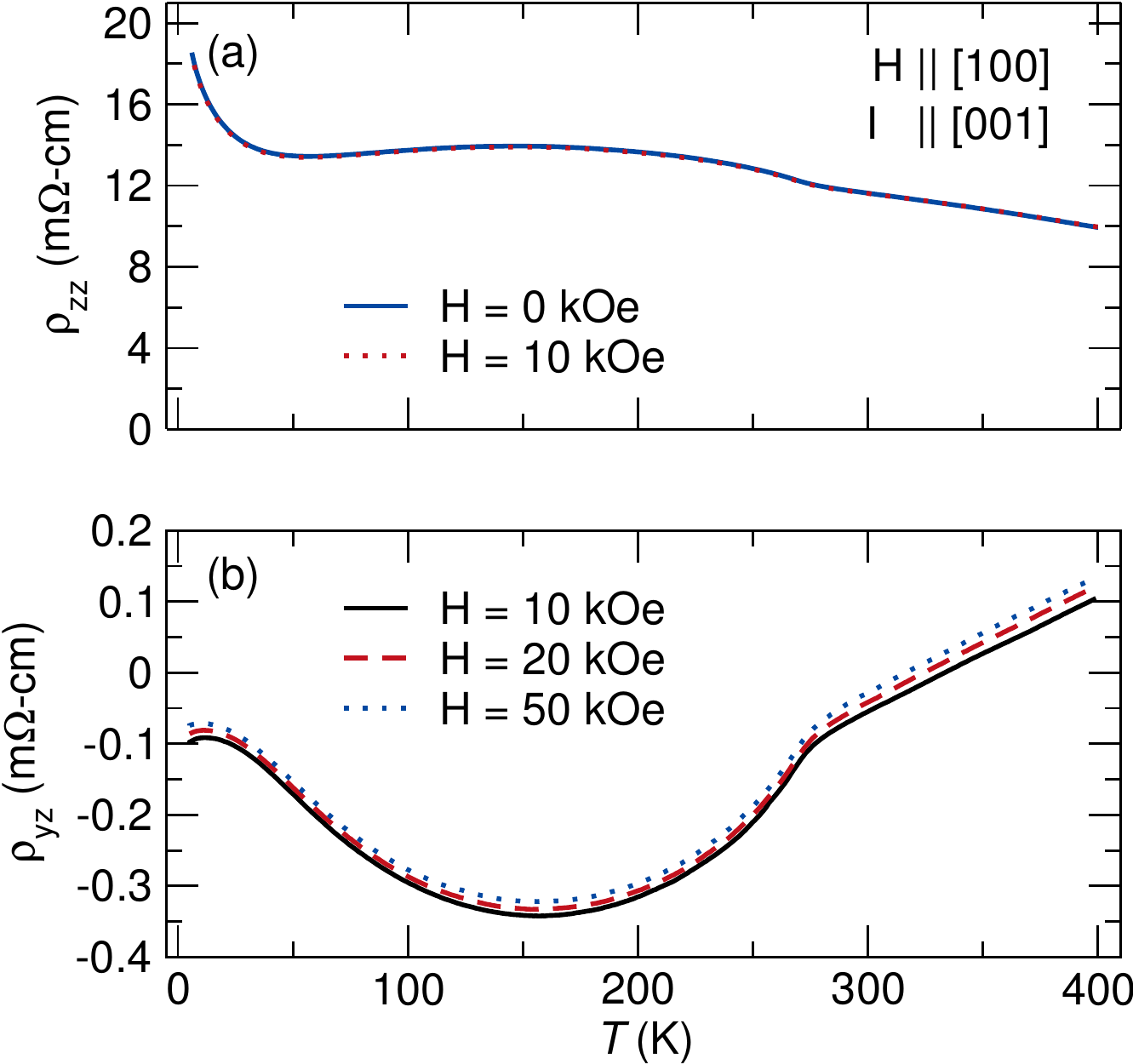} \\
\caption{\label{fig:res-data}(Color online.)
(a) Resistivity of \cumnas\ with applied field $H$ along [100] and current $I$ along [001]. The resistivity is relatively flat across the temperature range, with a small kink at $T_N = 270$~K. Hall measurements of the sample with $H$ along [100] and current along [001] show a decreasing trend followed by an increase at higher temperatures. 
}
\end{figure}

Four-point probe resistivity measurements along [001] show a mostly flat, weakly undulating trend versus temperature as shown in Fig.\ \ref{fig:res-data}(a).
The broad hump between 50~K to 250~K could be attributed to competing mobilities and carrier concentrations of multiple excited states in a heavily doped semiconductor (as in P-doped Si),\cite{Chapman1963} or variations in the dominant carrier scatterers in a disordered metal (which we discuss subsequently to be more likely, given the computed band structure).
The resistivity values are roughly 125 times higher at 5~K in \cumnas\ than Fe$_2$As and about 380 times higher than tetragonal CuMnAs, both of which are metallic \cite{Takeshita2017,Wadley2013}.
Application of a magnetic field of 10 kOe along [100]  resulted in negligible change in resistivity values, shown in Fig.\ \ref{fig:res-data}(a).
A slightly larger effect can be seen in the Hall effect measurements in Fig.\ \ref{fig:res-data}(b).
The Hall data magnifies the hump around 150~K, and crosses from negative (majority $n$-type) to positive ($p$-type) upon heating past 330~K. 
The material is $n$-type at low temperatures but as temperature is increased, more carriers are excited and the higher mobility of holes leads to compensation and switching to $p$-type conduction 330~K.
The lack of an anomaly in the total resistivity around the Hall crossover point indicates that the transport in \cumnas\ occurs via multiple bands, and is supported by the delicate (but not gapped) band structure around the Fermi energy  that we discuss subsequently.

\subsection{First-principles simulations}

\begin{figure}
\centering\includegraphics[width=\columnwidth]{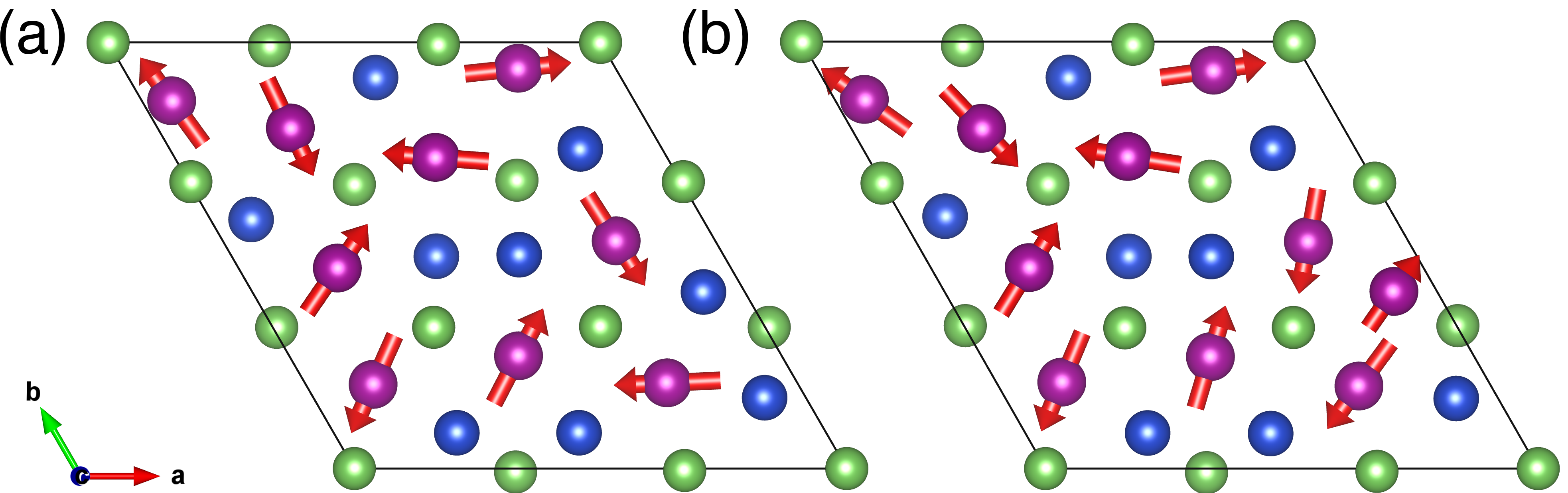} \\
\caption{\label{fig:DFT-mag}(Color online.)
Structure and magnetic configuration from DFT for (a) stoichiometric hexagonal CuMnAs and (b) Cu$_{0.89}$Mn$_{1.11}$As. 
Mn is shown in purple, Cu in blue, and As in green.
}
\end{figure}

We performed first-principles density-functional theory (DFT) simulations to confirm the stability, cell geometry, and magnetic
ordering of a fully-occupied hexagonal model compound CuMnAs and off-stoichiometric Cu$_{0.89}$Mn$_{1.11}$As, with a single Mn on a Cu1 site (1 of the 9 sites substituted per cell).
We find that the relaxed atomic geometries of hexagonal CuMnAs and Cu$_{0.89}$Mn$_{1.11}$As agree with neutron scattering results within $2\,\%$.
The DFT data for the magnetic structures arrive at different lowest-energy orderings than the neutron refinement. 
The DFT-derived lowest-energy magnetic configurations
of stoichiometric CuMnAs and substituted Cu$_{0.89}$Mn$_{1.11}$As are shown in Figs.\ \ref{fig:DFT-mag}(a) and (b), respectively.
The stoichiometric result is antiferromagnetic, while 
the substituted site in Cu$_{0.89}$Mn$_{1.11}$As has a small uncompensated moment ($-0.102a - 0.010b$~\textmu$_\mathrm{B}$). 
The calculated neutron diffraction structure factors for the stoichiometric case  are compared
to the single-crystal neutron-refined values
in Fig.\ \ref{fig:hb3a}(b). 
The two fits are similar, apart from  the trio of peaks with $F_{obs}^2 \approx 100$,
which significantly degrade the fit versus the neutron result.
The neutron refinement outperforms the DFT fit with $R_{F^2} = 7.77$ and $R_{F^2w} = 17.1$ versus $R_{F^2} = 7.98$ and $R_{F^2w} =23.0$, respectively, where smaller numbers indicate a better fit. 
{\color{black}A small uncompensated moment observed in the Mn-substituted DFT model is an unavoidable artifact of the cell choice, which contains one ``extra'' Mn atom to reflect the off-stoichiometry of \cumnas.}

Magnetic ground states in strongly-correlated $d$-electron systems are often challenging to predict using DFT,
so it is instructive to quantitatively evaluate the proximity of the neutron-refined result to the DFT energy minimum, and likewise the predicted neutron intensities of the DFT model.
To better understand the energetics of this difference between theory and experiment, we compare total energies for three different situations:
First, chemical and magnetic structures are constrained to the neutron scattering result ($E_\mathrm{fix}$ in Table \ref{tab:DFT-energy-latparam}).
Second, the ground-state magnetic structure is computed from DFT while the atomic geometries are constrained to the neutron scattering data ($E_\mathrm{mag}$ in Table \ref{tab:DFT-energy-latparam}).
Finally, these total energies are compared to the fully relaxed DFT result ($E_\mathrm{all}$ in Table \ref{tab:DFT-energy-latparam}).
These small energy changes, 15.30 and 14.93~meV/atom for CuMnAs and Cu$_{0.89}$Mn$_{1.11}$As, respectively are typical of energy differences between various magnetic structures for similar systems.\cite{Alsolami2012Auth}

\begin{table}
\caption{\label{tab:DFT-energy-latparam} 
Energy differences (meV/atom) between different constraints in DFT (see text), and lattice parameters (\AA, degree) from all-relaxed calculations of stoichiometric hexagonal CuMnAs and Cu$_{0.89}$Mn$_{1.11}$As. All phases have $\gamma = 120^\circ$.
}
\centering
\begin{tabular}{cccccc}
\hline\hline
System	 & $E_\mathrm{fix}-E_\mathrm{mag}$	 & $E_\mathrm{fix}-E_\mathrm{all}$	 & $a$	 & $b$	 & $c$ \\
\hline
CuMnAs	 & 9.92	 & 15.30	 & 11.050	 & 11.050	 & 3.802\\
Cu$_{0.89}$Mn$_{1.11}$As	 & 6.45	 & 14.93	 & 11.053	 & 11.043	 & 3.776\\
\hline \hline
\end{tabular}
~\\
\end{table}




\begin{figure}
\centering\includegraphics[width=\columnwidth]{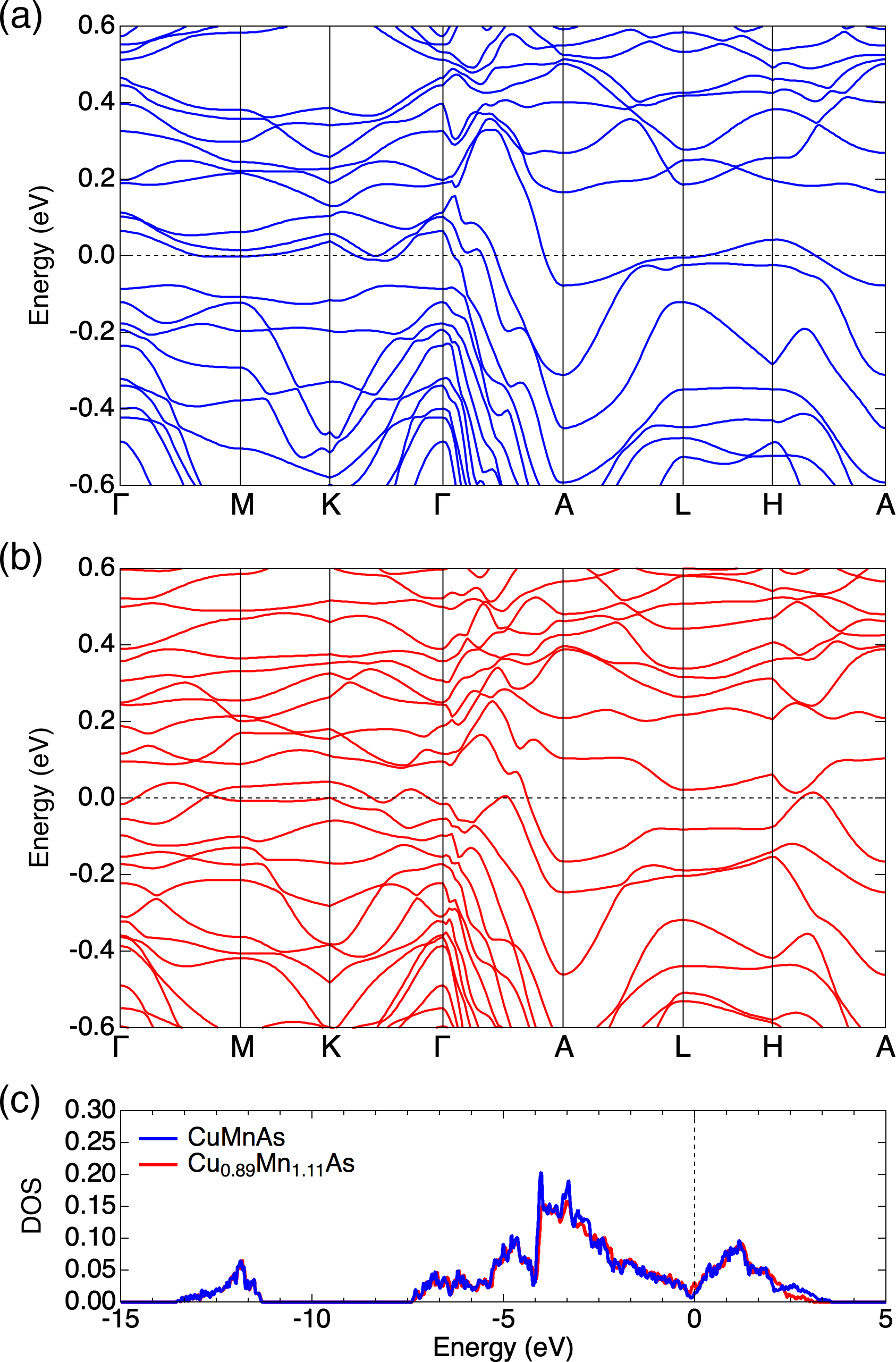} \\
\caption{\label{fig:DFT-band}(Color online.)
Electronic band structure of (a) stoichiometric hexagonal CuMnAs (blue) and (b) Cu$_{0.89}$Mn$_{1.11}$As (red). Both densities of states (DOS, in units of states per \AA$^3$ and per eV per spin), computed using DFT, are shown in (c).
The highest-occupied energies are set as $E=0$ eV.
}
\end{figure}

The electronic band structure and density of states of stoichiometric hexagonal CuMnAs and Cu$_{0.89}$Mn$_{1.11}$As in Fig.\ \ref{fig:DFT-band} show that both hexagonal models are metallic.
Both electronic structures exhibit very small densities of states near the Fermi energy, similar to that described by DFT for tetragonal CuMnAs \cite{Maca2017}.
Tetragonal CuMnAs shows obvious metallic resistivity ($d\rho/dT > 0$).\cite{Wadley2013}
The CuMnAs compounds are clearly on the cusp of semiconducting/metallic behavior, and share similarities to Fe$_2$As, which has a much greater density of states at the Fermi level and does show $d\rho/dt > 0$, but the reported values of resistivity values are much higher than that of tetragonal CuMnAs.\cite{Yang2019,Takeshita2017}

Our DFT calculations suggest that Cu$_{0.89}$Mn$_{1.11}$As is metallic. However, transport measurements indicate that the resistivity is high, and $d\rho/dT < 0$ for most $T$. 
The negative slope that is observed at low and high temperatures is not exponential as is expected in highly-doped semiconductors.\cite{Chapman1963}
The seeming discrepancy between resistivity and the computed band structure can be resolved by considering the high amount of substitutional disorder in these compounds.
Metals often exhibit $d\rho/dT < 0$ behavior when a large amount of configurational disorder is present,\cite{Elk1979} and the negative temperature dependence is in fact correlated with high absolute values of resistivity.\cite{Mooij1973} 
In our material, carriers must scatter due to pervasive disorder due to Mn site mixing, while magnon scattering may also contribute strongly, but the overall resistivity is hardly affected upon cooling past $T_N$.

\section{Conclusions}

We report the crystal structure of a non-centrosymmetric $P\overline{6}^\prime$ phase in the Cu--Mn--As system, with a new structure type. This compound can be made phase-pure in single crystal form. Triangular antiferromagnetic ordering appears upon cooling below 270~K and is markedly distinct from the orthorhombic and tetragonal CuMnAs phases, both of which are stabilized by different Cu/Mn content and are centrosymmetric in their paramagnetic states. DFT calculations confirm the stability of the magnetic structure refined by single-crystal neutron diffraction. The triangular AF ordering is in-plane and does not break degeneracy of the $a$ and $b$ axes.  Like other copper manganese arsenides, hexagonal \cumnas\ is on the cusp of semiconducting/metallic behavior and further investigation of the carrier scattering mechanisms in this class of materials is warranted.

\section{Acknowledgments}

This work is supported by the National Science Foundation
(NSF) under Grant No.\ DMR-1720633.
Characterization was carried out in part in the Materials Research Laboratory Central Research Facilities, University of Illinois. Use of the Advanced Photon Source at Argonne National Laboratory was supported by the U.S.\ Department of Energy, Office of Science, Office of Basic Energy Sciences, under Contract No.\ DE-AC02-06CH11357. Neutron scattering was performed at the High Flux Isotope Reactor, a Department of Energy Office of Science User Facility operated by the Oak Ridge National Laboratory.
This work made use of the Illinois Campus Cluster, a computing resource that is operated by the Illinois Campus Cluster Program (ICCP) in conjunction with the National Center for Supercomputing Applications (NCSA) and which is supported by funds from the University of Illinois at Urbana-Champaign.
We thank Junseok Oh for assistance in making contacts for resistivity measurements.
\bibliography{h-cumnas}

\end{document}